\documentclass[fleqn,usenatbib]{mnras}

\usepackage{newtxtext,newtxmath}

\usepackage[T1]{fontenc}
\usepackage{ae,aecompl}

\usepackage{graphicx}	
\usepackage{amsmath}	
\usepackage{amssymb}	

\title[\textit{K2} observations of PSR J1023$+$0038]{\textit{Kepler} \textit{K2} observations of the transitional millisecond pulsar PSR J1023+0038}

\author[M. R. Kennedy et al.]{
M. R. Kennedy$^{1}$\thanks{E-mail: kennedy.mark@manchester.ac.uk}\thanks{These authors contributed equally to this work.},
C. J. Clark$^{1}$\footnotemark[2],
G. Voisin$^{1}$\footnotemark[2],
R. P. Breton$^{1}$
\\
$^{1}$Jodrell Bank Centre for Astrophysics, School of Physics and Astronomy, The University of Manchester, M13 9PL, UK
}

\date{Accepted 2018 March 14. Received 2018 March 12; in original form 2018 January 31}

\pubyear{2018}

\begin{document}
\label{firstpage}
\pagerange{\pageref{firstpage}--\pageref{lastpage}}
\maketitle

\begin{abstract}
For 80 days in 2017, the \textit{Kepler Space Telescope} continuously observed the transitional millisecond pulsar system PSR J1023$+$0038 in its accreting state. We present analyses of the 59-second cadence data, focusing on investigations of the orbital light curve of the irradiated companion star, and of flaring activity in the neutron star's accretion disc. The underlying orbital modulation from the companion star retains a similar amplitude and asymmetric heating profile as seen in previous photometric observations of the system in its radio pulsar state, suggesting that the heating mechanism has not been affected by the state change. We also find tentative evidence that this asymmetry may vary with time. The light curve also exhibits ``flickering'' activity, evident as short time-scale flux correlations throughout the observations, and periods of rapid mode-switching activity on time scales shorter than the observation cadence. Finally, the system spent $\sim 20\%$ of the observations in a flaring state, with the length of these flares varying from $<2$~minutes up to several hours. The flaring behaviour is consistent with a self-organised criticality mechanism, most likely related to the build up and release of mass at the inner edge of the accretion disc.
\end{abstract}

\begin{keywords}
pulsars: individual (PSR J1023$+$0038) -- accretion, accretion discs -- X-rays: binaries -- stars: neutron -- binaries: general -- instabilities
\end{keywords}



\section{Introduction}

``Redback'' binary star systems contain a rotating neutron star primary, and a semi-degenerate companion star with a mass of $0.1$--$0.5$~M$_{\odot}$ in a close orbit. The neutron star (NS) rotates with a millisecond period, and is thought to be the descendant of a low-mass X-ray binary (LMXB), having been spun-up by accreting matter from the companion (\citealt{1982Natur.300..728A}; \citealt{1982CSci...51.1096R}). These systems take their name from the ``redback'' spider species, in which the female consumes the male spider after mating.  For an overview of ``redbacks'' and their two states, and of their close cousins, the ``black widow'' systems (with lower-mass $<0.05$~M$_{\odot}$ companions), see \citet{2013IAUS..291..127R}.

In recent years, three redback systems have become increasingly important in understanding the binary origin of millisecond pulsars (MSPs). These three systems, PSR J1023+0038 \citep{2009Sci...324.1411A}, PSR J1824$-$2452I \citep{2013Natur.501..517P} and PSR~J1227$-$4853 \citep{2014MNRAS.441.1825B,2015ApJ...800L..12R}, make up the class known as ``transitional'' millisecond pulsars (tMSPs). tMSPs are binary systems which have been observed to switch between two very distinct states -- a rotation powered state, and an accretion powered state. In the rotation powered states, these systems have high radio luminosities and the NS primary is detectable through its radio pulsations as an MSP. In the accretion powered state, no pulsations have been detected at radio wavelengths, but the systems appear far brighter at optical, X-ray and (in the cases of PSRs J1023$+$0038 and J1227$-$4853) gamma-ray wavelengths \citep{2014ApJ...790...39S,2017ApJ...836...68T}, interpreted as the sudden appearance of an accretion disc in the system.

PSR J1023+0038 (hereafter J1023) was the first member of the redback class of tMSPs to be discovered. It was originally classified as a cataclysmic variable (a binary star system with a white dwarf primary) in 2001 after the radio source FIRST J102347.6+003841 was associated with a Galactic optical counterpart which had a very blue optical spectrum, and which also showed double peaked emission lines which were associated with an accretion disc (\citealt{2002PASP..114.1359B}; \citealt{2003AJ....126.1499S}). Further optical photometry was taken in 2003 \citep{2004MNRAS.351.1015W}, but the light curve lacked the large amplitude flickering events which dominated the light curve in 2001. \citet{2005AJ....130..759T} confirmed a state transition, as the optical spectrum was then dominated by strong absorption features and lacked any prominent emission components, and further suggested that the system was not a cataclysmic variable, but could harbour a neutron star binary instead. 

In 2007 the system was detected as a radio pulsar with a spin period of 1.69 ms \citep{2009Sci...324.1411A}. J1023 was reclassified as an MSP binary system. This was the first direct evidence of an MSP in a binary system transitioning between two distinct states, and gave significant support to the ``recycled'' scenario for the origin of MSPs. This scenario was confirmed in 2013, when radio pulsations from the system became undetectable, and the gamma ray luminosity of the system increased dramatically \citep{2013ATel.5513....1S,2014ApJ...790...39S}. Follow up observations at various wavelengths confirmed that the radio pulsar had switched off, the X-ray and UV flux of the system had increased, and an accretion disc had begun reforming in the system \citep{2014ApJ...785..131T}.

Since then, the system has exclusively occupied the accretion powered state. However, the system has shown dramatic variability in this state. At X-ray wavelengths, the system displays three modes: a low mode (during which X-ray pulsations are not detectable), a high mode (during which X-ray pulsations are detectable), and a flaring mode \citep{2015ApJ...806..148B}. The duration of these modes is highly variable, and it has also been recently confirmed that the radio flux from the system is anti-correlated with the X-ray flux (that is, during the high X-ray mode, the radio flux is at a minimum, and during the X-ray low mode, the radio flux reaches a maximum; \citealt{2017arXiv170908574B}). There are no reported detections of radio pulsations from the system in the accretion state, but there is persistent radio emission which has a flat spectrum. This radio emission has been suggested to originate from synchrotron radiation from a partially self-absorbed jet \citep{2015ApJ...809...13D}. At optical wavelengths, there also appears to be three modes: a low mode, a high mode, and a flaring mode (\citealt{2015MNRAS.451.3468M}; \citealt{2015MNRAS.453.3461S}). This is best seen in several of the optical light curves presented in \citet{2015MNRAS.453.3461S}, which show very high amplitude optical flares alongside a much lower bimodal distribution in optical flux. It should be noted that this bimodal distribution is not always visible in optical data. 

A common physical scenario has been proposed to explain these modes and why the system switches between them so often. During the high X-ray state, the accretion disc is thought to be truncated, and material flows from the disc towards the NS, generating X-rays which are modulated at the spin period of the NS, and radio emission through the suggested jet is quenched. During the X-ray low mode, material does not couple from the accretion disc onto the NS, but rather just builds up in the inner part of the disc. This leads to a lower X-ray flux, and the radio emission from the jet is no longer quenched, increasing the radio flux. Finally, the material which builds up at the inner edge of the disc is eventually ejected from the inner regions of the system by the rotating magnetic field of the NS in what has been dubbed a ``propeller'' state \citep{2015ApJ...807...33P}. The existence of a propeller state in J1023 has recently gained support from both optical polarimetry measurements and Doppler tomograms \citep{2018MNRAS.474.3297H}. This state has been used to great effect in explaining the emission in some magnetic cataclysmic variables (e.g. AE Aqr; \citealt{1997MNRAS.286..436W}). Interactions between the pulsar's co-rotating magnetosphere and the inner regions of the accretion disc have also been proposed to explain the extremely high luminosity of optical pulsations discovered recently at the spin period of the MSP in the J1023 system \citep{2017NatAs...1..854A}.

Here we present \textit{Kepler} \textit{K2} \citep{2014PASP..126..398H} observations of J1023, the first ever long term continuous optical observations of a redback system. We first investigate the orbital modulation in the light curve by fitting two sinusoidal components at the orbital frequency and its second harmonic (Section~\ref{sec:Timing}). We then subtract this model from the full data, and investigate the flickering (Section~\ref{sec:Flickering}) and flaring (Section~\ref{sec:Flares}). Finally, we discuss these results in the context of previous observations of J1023, and propose a self-organised criticality origin for the flaring activity in Section~\ref{sec:Discussion}. 

At a late stage in our investigation of these data, a similar study was presented in \citet[][hereafter P18]{2018arXiv180104736P}. Our conclusions are in broad agreement, with some notable exceptions which will be addressed at the relevant points in our manuscript. 
\defcitealias{2018arXiv180104736P}{P18}

\section{Observations and Reduction}\label{sec:reduction}

J1023 was observed by the \textit{Kepler} spacecraft as part of Campaign 14 of the \textit{K2} mission from 2017 May 31 to 2017 August 19, with both long cadence ($\sim$30 min) and short cadence ($\sim$1 min) observations recorded. The \textit{Kepler} bandpass is broad, covering 4200 \AA\ to 9000 \AA, and peaking at 6000 \AA. The reduced long cadence light curve was downloaded from MAST, while the short cadence target pixel file (TPF) was downloaded, and a light curve extracted using a custom aperture. The mask used to define the aperture was basic and covered all pixels close to the target, as J1023 was the only object in the TPF.

The light curve was filtered by removing data points where the error flag in the TPF was $>0$ (such points correspond to observations taken when the spacecraft was firing its thrusters which occurs every $\sim$6 hours to $\sim$ 2 days).

\begin{figure*}
	\includegraphics[width=0.99\textwidth]{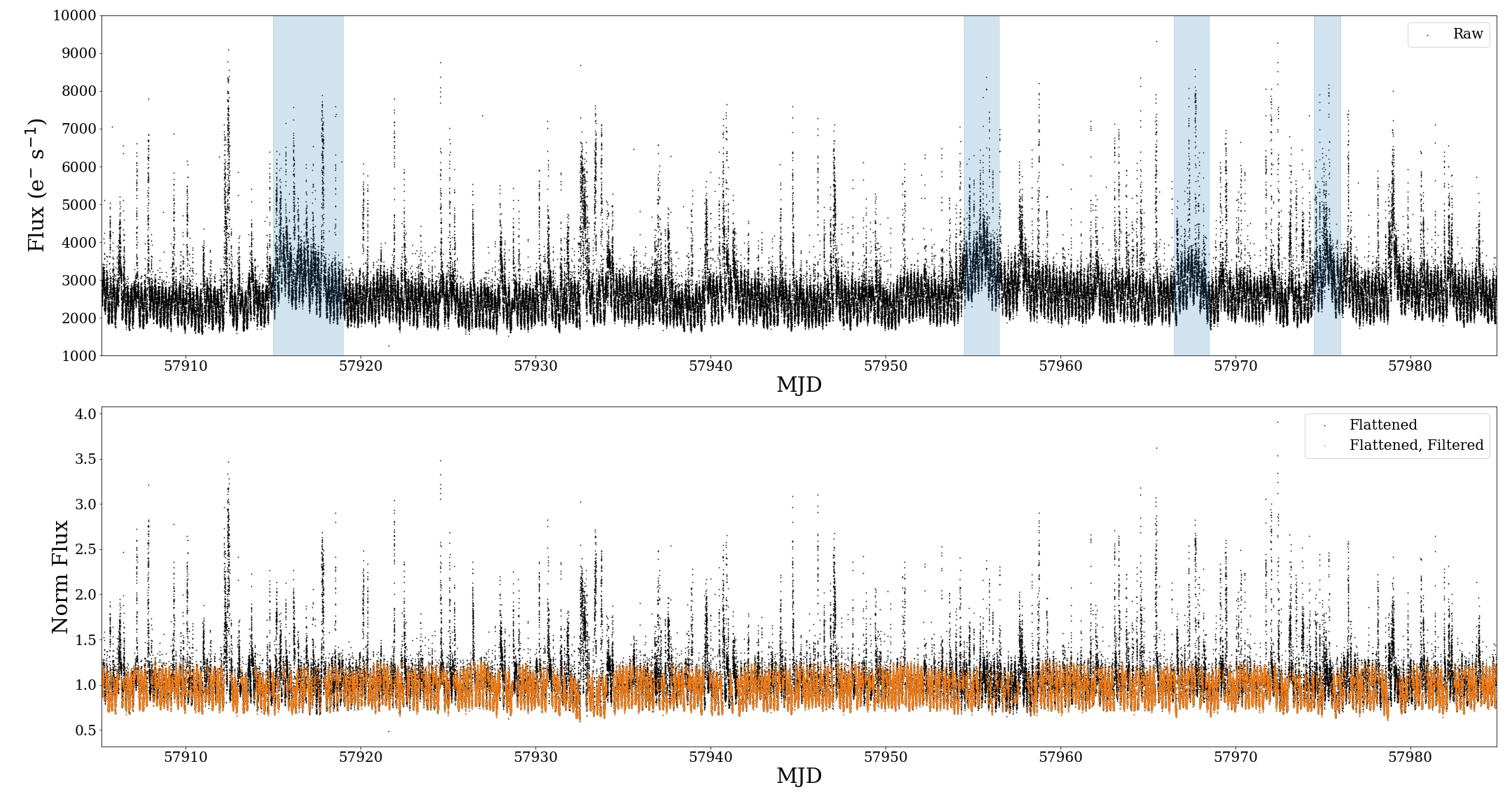}
    \caption{\textit{Top}: The extracted 80 day light curve of J1023. The light curve shows flaring events, as well as several periods of increased median brightness (highlighted). The cause of this enhanced brightness is not yet known. \textit{Bottom}: The flattened light curve (black) and the sigma clipped light curve (orange) which was used in the orbital modulation analysis.}
    \label{fig:full_lc}
\end{figure*}

To ensure the quality of the short cadence light curve, we compared the extracted short cadence light curve to the long cadence light curve obtained using the Pre-search Data Conditioning Simple Aperture Photometry (PDCSAP; \citealt{2012PASP..124.1000S}) pipeline to look for any effects introduced from extraction using our custom mask. The comparison showed the same long term behaviour in both the short and long term light curves. The extracted light curve is shown in the top panel of Figure~\ref{fig:full_lc}. Finally, we created a flattened light curve by removing the low frequency variability using the \textsc{Pyke} task \textsc{kepflatten} (\citealt{2012ascl.soft08004S}; \citealt{pyke3}). \textsc{kepflatten} works by fitting a polynomial to windows of data of a specified length, after outlier points have been removed, and then combining these polynomials together and dividing the data by the resulting piecewise function. Our parameters for flattening the data were \textsc{nsig} $=3$ ($\sigma$ clipping threshold), a step size of 1 day, a window size of 3 days, a polynomial of order 3, and the process was iterated 10 times.

\section{Timing Analysis}\label{sec:Timing}

\begin{figure}
	\includegraphics[width=\columnwidth]{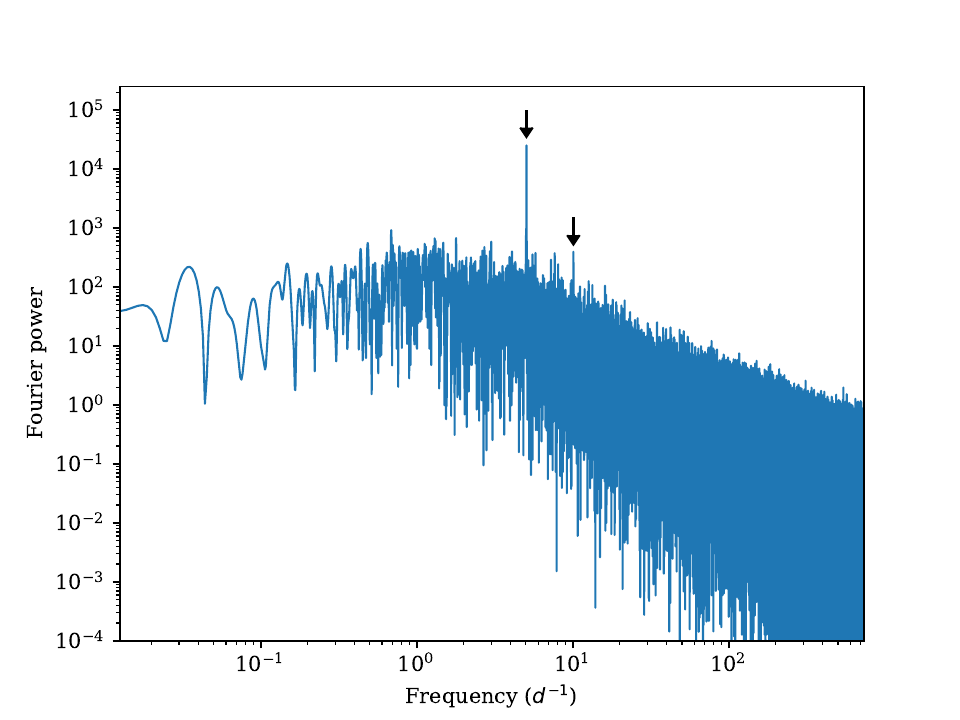}
    \caption{Power spectrum of the flattened data calculated via Fourier Transform normalised such that the spectrum of white Gaussian noise would have a $\chi^2$ distribution with two degrees of freedom. The power spectrum shows strong power at the orbital frequency and its second harmonic, highlighted by arrows. The intrinsic pink noise of the light curve is visible as a descending power-law at frequencies above $1$~d$^{-1}$, with the turn over below this frequency being due to the flattening performed in Section~\ref{sec:reduction}. }
    \label{fig:powspec}
\end{figure}

The power spectrum of the light curve (Figure~\ref{fig:powspec}) shows clear modulation at both the orbital frequency, $\nu_{1}$, and twice the orbital frequency, $\nu_{2} = 2\nu_{1}$ (hereafter referred to as the second harmonic). These signals are common to the optical light curves of many redback pulsars, and are attributed to the heating of the side of the companion facing the neutron star (producing a modulation at the orbital period with maximum at companion superior conjunction), and the ellipsoidal shape of the companion caused by tidal deformation in the strong gravitational potential of the neutron star (producing a modulation at half the orbital period, with maxima at ascending and descending nodes), respectively \citep[e.g.][]{2013ApJ...769..108B}. We hereafter refer to the signal at $\nu_1$ as the \textit{heating} component, and the signal at $\nu_2$ as the \textit{ellipsoidal} component, although we note that the true distinction between these components is not straightforward, as a non-sinusoidal signal at $\nu_1$ will also introduce power at $\nu_2$. 

These signals need not be perfectly in phase with the orbital motion of the system. Asymmetric heating will manifest as a phase shift between the heating and ellipsoidal components, and a time-varying asymmetry may be detectable as a slight difference in the periods of the two components. The long time span and high cadence of the \textit{Kepler} data allow us to investigate these effects. 

To compare our measured orbital phases and periods, we used a reference ephemeris for the system's orbit, derived from timing observations of the neutron star's X-ray pulsations. We first took the orbital period ($P = 0.19809664676$~d) and period derivative ($\dot{P} = -1.65(19)\times10^{-10}$) measured by \citet{2016ApJ...830..122J} with a reference epoch of MJD $54905.9694347$, and extrapolated the orbital period to the centre of the \textit{Kepler} observations. Using the resulting orbital period, and the time of ascending node given by \citetalias{2018arXiv180104736P}, $T_{\rm ASC} = 57896.8292633(2)$, derived from \textit{XMM-Newton} observations performed shortly before the \textit{Kepler} observations, we found an ascending node close to the centre of \textit{Kepler} data.  This results in a reference ephemeris with $T_{\rm ASC} = 57944.9666256(6)$ and $P = 0.198096145(2)$~d. The orbital period derivative measured by \citet{2016ApJ...830..122J} corresponds to a shift in the time of ascending node of less than one second between the start and centre of the \textit{Kepler} observations. We therefore assumed a constant orbital period in our analysis. 

As a first step, we computed the power spectrum of the light curve (shown in Figure~\ref{fig:powspec}) via a discrete Fourier transform. The power spectrum was interpolated by zero-padding the evenly sampled \textit{Kepler} time series to 10 times its original length. The two most significant peaks in the power spectrum are at the known orbital period of J1023, and its second harmonic. However, as can be clearly seen from Figure~\ref{fig:powspec}, the light curve shows considerable \textit{pink noise}: correlated noise between neighbouring observations with a descending power-law spectrum which follows a relation close to $1/f$. This pink noise will add power at the orbital frequency and its harmonics, biasing measurements of the period and phase of the orbital modulations and will be discussed in detail later. Flares additionally reduce the significance of the orbital modulations.

At this stage we are only interested in the orbital modulation of the companion star. We therefore attempted to remove flares from the data. To do this, we first subtracted an initial model for the orbital modulation of the form 
\begin{equation}\label{e:orbital_model}
f(t) = A_1 \cos(2\pi\,\nu_1 (t - t_0) + \phi_1) + A_2 \cos(2\pi\,\nu_2 (t - t_0) + \phi_2),
\end{equation}
where $t_0$ is a reference time at the centre of the \textit{Kepler} observations, chosen to remove correlations between frequencies $\nu_1,\nu_2$ and phases $\phi_1,\phi_2$. The initial amplitudes, frequencies, and phases were derived from the power spectrum of Figure~\ref{fig:powspec}. We then iteratively removed data points which lay more than 2.5 standard deviations above unity after dividing by a running median evaluated over $1$~day windows around each data point. To further reduce flare contamination, we additionally removed the 10 data points closest to each data point which lay above our threshold. This aggressive strategy removed $\sim40\%$ of data points, leaving a set of ``clean'' points dominated by the orbital modulation, with pink noise ``flickering'' remaining throughout the data. This data is shown in orange in Figure~\ref{fig:full_lc}.

To partially account for the effects of this pink noise we adopt the ``whitening'' methods used by \citet{2011MNRAS.418..561C} to measure the rotational properties of radio pulsars in the presence of strong red noise. The goal of these methods is to transform the data into a new basis in which the measured residuals have a frequency-independent (``white'') normalised Gaussian distribution. The best-fitting orbital model is found by minimising the chi-squared statistic,
\begin{equation}
  \chi^2 = \vec{R}^T C^{-1} \vec{R}\,.
  \label{e:generalised_least_squares}
\end{equation}
where $\vec{R}$ is the vector of residuals (i.e. the data with the model subtracted) and $C$ is the covariance matrix.

To estimate the covariance matrix, we computed the average covariance function, $c(\tau) = \langle f(t) f(t+\tau) \rangle$, where angle brackets denote averages over pairs of data points with the same lag, $\tau$. To avoid removing signals at the orbital frequency, which would appear as periodicities in the covariance function, we fit a smooth function consisting of three decaying exponential terms to the average covariances, i.e.,
\begin{equation}
  \hat{c}(\tau) = \sum_{i}^3 k_i\, \mathrm{e}^{-\tau/\lambda_i}\,,
  \label{e:covariance_matrix}
\end{equation}
where $k_i$ and $\lambda_i$ are amplitude and scale parameters. The best-fitting values for these parameters are given in Table~\ref{tab:covar}.

\begin{table}
	{\centering
	\caption{\label{tab:covar} Parameters of the covariance function, Equation~\ref{e:covariance_matrix}}
	\begin{tabular}{r c c}
		\hline
		Component, $i$ & $\lambda_i$ (d) & $k_i$\\
		\hline\hline
        1 & $6.75\times10^{-4}$ & $1.88\times10^{-3}$\\   
        2 & $5.94\times10^{-2}$ & $8.52\times10^{-4}$\\   
        3 & $8.07\times10^{-1}$ & $6.57\times10^{-4}$\\   
        \hline
	\end{tabular}}
\end{table}

In practice, the full covariance matrix is too large to feasibly work with. Fortunately, an effect of the flattening performed in Section~\ref{sec:reduction} is to remove covariances longer than the window time scale (3 days, approximately 15 orbits), and the covariance function drops to zero for longer lags. The covariance matrix therefore has a ``banded'' form, with all elements far from the diagonal being zero. To ensure the covariance matrix remained positive definite, we further multiplied the covariance function by a Hann (a cosine) window dropping to zero after four days \citep[as in][]{2015ApJ...812..128C}. With only the non-zero elements of the covariance matrix stored in this banded form, we make use of efficient functions implemented by the \texttt{scipy}\footnote{\url{https://www.scipy.org/}} module to calculate Equation \ref{e:generalised_least_squares}. As in \citet{2011MNRAS.418..561C} we also make use of the Cholesky decomposition $C = L L^T$, in which $L$ is a lower-triangular matrix, to further speed up the computation. The periods, phases and amplitudes of the fundamental orbital modulation and its second harmonic are then fit by minimising Equation~\ref{e:generalised_least_squares}. The resulting best-fitting values are given in Table~\ref{tab:measured_ephem}.

\begin{table}
	{\centering
	\caption{\label{tab:measured_ephem}
Orbital light curve fitting results}
	\begin{tabular}{r c c}
		\hline
		Parameter & Reference Ephemeris	& Best-fitting offset\\
		\hline\hline
        $T_{\rm ASC}$ (MJD) & $57944.9666256(6)$ & ---\\
        $T_{{\rm ASC},1}$ (MJD)$^{a}$ & --- & $-5.2(2)\times10^{-3}$ \\
        $T_{{\rm ASC},2}$ (MJD)$^{a}$ & --- & $3.2(5)\times10^{-3}$ \\
        \hline
        $\nu_1$ (d$^{-1}$)$^{b}$ & $5.04805380(6)$ & $2.0(4)\times10^{-4}$ \\
        $\nu_2$ (d$^{-1}$)$^{b}$ & $10.0961076(1)$ & $-7(2)\times10^{-4}$ \\
        \hline
        $A_1$$^{c}$ & --- & $0.163(1)$ \\
        $A_2$$^{c}$ & --- & $0.02(1)$ \\
        \hline
	\end{tabular}}
    Notes --- Best-fitting offsets (third column) between the reference ephemeris (second column, see text) and the heating (subscript $1$) and ellipsoidal (subscript $2$) components of our model (Equation \ref{e:orbital_model}), with $1\sigma$ uncertainties on the final digits in brackets.\\
    $^{a}$ Inferred ascending node epochs of the components of our model\\
    $^{b}$ Frequencies of the model components, expected to be equal to the orbital frequency and its second harmonic.\\
    $^{c}$ Amplitudes of the model components expressed as fractions of the average flux.
\end{table}

The Cholesky decomposition additionally allows us to investigate the ``whitened'' residuals, $\vec{R}_{\rm W} = L^{-1} \vec{R}$, which should have a normalised Gaussian distribution. In practice, we find that the distribution of $\vec{R}_{\rm W}$ is narrower around the peak and wider in the tails than a true Gaussian. We attribute this to time-varying covariances in the data caused by differing activity modes in the accretion disc (see Section \ref{sec:Flickering}), which are not captured by our stationary covariance function. To partially address this, we down-weighted periods of higher activity by multiplying each row and column of the covariance matrix by the running variance of the whitened residuals. Despite this, the distribution of whitened residuals remains clearly non-Gaussian. We also find that the residual distribution is not symmetric, but peaks slightly below zero. We believe that this is due to some remaining flaring activity from the accretion disk, which will always increase the observed flux, extending the positive tail of the residual distribution (see Section~\ref{sec:Flares} for more details).

\begin{figure}
	\includegraphics[width=\columnwidth]{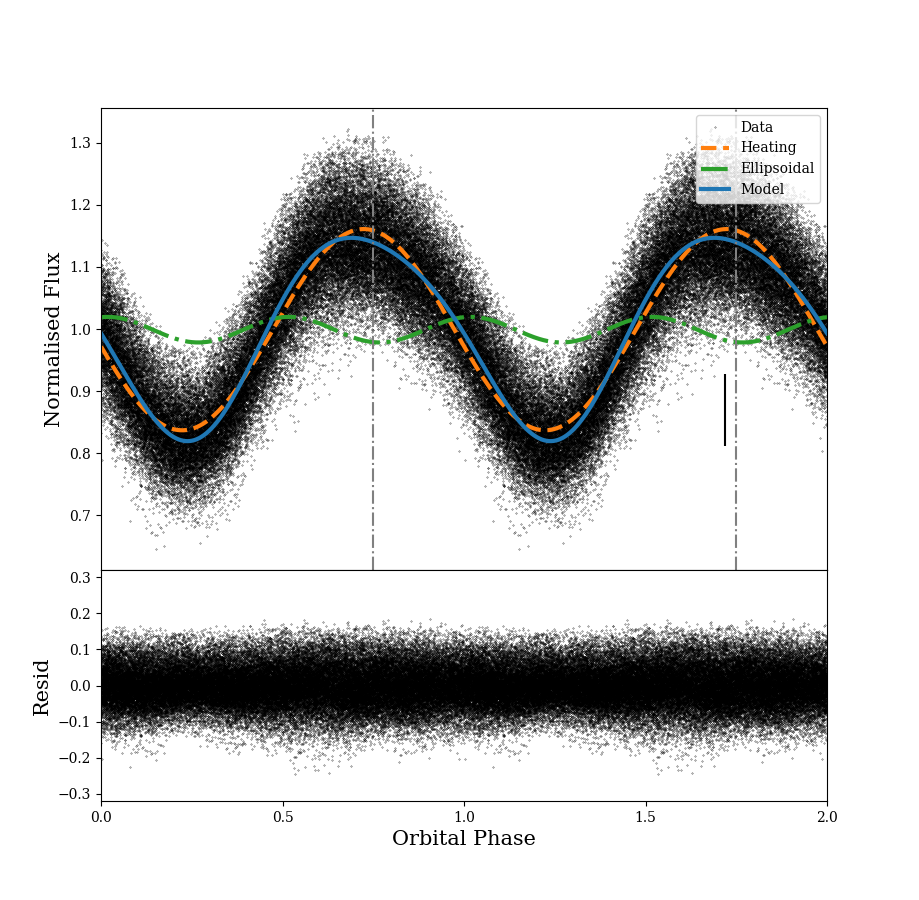}
    \caption{Orbital light curve of the companion star. The black points show the short-cadence \textit{Kepler} data, flattened over 3-day windows, after subtracting data points around flares, and normalising by the median. The best-fitting model is shown by the solid blue curve, the ``heating'' component to the best-fitting curve is shown by the dashed orange line, and the ``ellipsoidal'' component by the green dot-dashed line. An orbital phase of zero corresponds to the neutron star's ascending node, determined by (unpublished) measurements of the Doppler shift of X-ray pulsations by \citetalias{2018arXiv180104736P}. We note that the maximum is asymmetric, and occurs prior to the companion superior conjunction at phase 0.75 which is marked with a grey dashed vertical line. A single error bar is shown at phase 1.72, the value of this error was found as described in Section \ref{sec:Model}. Two identical periods are shown for clarity.}
    \label{fig:model_lc}
\end{figure}

The best-fitting model to the orbital light curve is shown in Figure~\ref{fig:model_lc}. We find that the heating component (i.e. the sinusoidal modulation closest to the orbital frequency) arrives earlier than expected from the pulsation-derived ephemeris by 7.6$\pm$0.3~minutes, while the ellipsoidal component (at the second harmonic of the orbital frequency) arrives late by 4.5$\pm$0.7~minutes with respect to the orbital ephemeris. The result is a distinctly asymmetric maximum with a sharper rise and shallower fall than a perfect sinusoid, peaking before companion superior conjunction. The measured amplitudes and phases of the two components agree with the values obtained by \citetalias{2018arXiv180104736P} by averaging over multiple 2-day windows.

The best-fitting orbital frequencies $\hat{\nu}_{1}$ and $\hat{\nu}_2$ are inconsistent with the X-ray pulsation timing-derived ephemeris period, appearing faster and slower at the $5\sigma$ and $3\sigma$ levels respectively. The difference in $\nu_1$ corresponds to a time offset in the maximum of the heating component of $\pm2.3(5)$~minutes at the start/end of the \textit{Kepler} observations with the maxima becoming more asymmetric over the course of the observations. However, as we were unable to completely remove the correlated noise from the light curve, we note that the best-fitting periods of the two components and their statistical uncertainties should be treated with caution. 

We also repeated the fitting procedure while allowing the amplitudes of the two components to vary linearly with time. No significant variation was detected, with 2$\sigma$ upper limits on the fractional amplitude derivatives of $\left|\dot{A}_1\right| < 8\times10^{-5}$~d$^{-1}$ and $\left|\dot{A}_2\right| < 5\times10^{-5}$~d$^{-1}$ respectively.

\subsection{Monte Carlo Analysis}\label{sec:Model}

To confirm the asymmetries in the orbital light curve and the values arrived at in the previous section, we took the model described by Equation \ref{e:orbital_model} and estimated the errors on the parameters using a Monte Carlo Markov Chain (MCMC) with top hat priors on all of the parameters. The data was first folded on the known orbital period using the reference ephemeris given in Table~\ref{tab:measured_ephem}, and the period was not allowed to vary in our MCMC analysis. Additionally, the predicted error on any data point from the \textit{K2} mission for an object with a V band magnitude of $\sim 16.5$ taken in short cadence mode is $\sim 1\%$ \citep{2014PASP..126..948V}. To establish if this estimated error is appropriate for the data presented here, we also allowed the error ($\sigma$) on the data points to vary in our MCMC fits (we also assumed that the error on every data point was the same). For this parameter, we assumed a Jeffreys prior (that is, $P(\sigma) \propto 1/\sigma$). The priors are also given in Table~\ref{tab:model_lc_results}. We also included a parameter to allow the baseline of the model to vary from 1 ($A_{0}$). 

The MCMC was carried out using \textsc{emcee} \citep{2013PASP..125..306F}. A total of 50 walkers were used with 25,000 steps taken and a burn in of 5000 steps allowed. Inspection of the posteriors of the parameters did not show any obvious correlations between any of the parameters and the individual parameter distributions followed normal distributions.

\begin{table}
	\centering
	\caption{Results from fitting harmonic series to light curve. The numbers underneath $\phi_1$ and $\phi_2$ represent the phase offsets from the expected arrival phases of 0.5 and 0.25, respectively, in minutes.}
	\begin{tabular}{r c c}
		\hline
		Parameter		& Prior				& Value\\
		\hline\hline
        $A_{0}$			& $0.8<A_{0}<1.1$		& $1.0001\pm0.0005$\\
        $A_{1}$			& $0.1<A_{1}<0.3$		& $0.1615\pm0.0009$\\	
        $A_{2}$			& $0.005<A_{2}<0.1$ 	& $0.020\pm0.008$\\
        $\phi_1$		& $0.0<\phi_1<1.0$ 		& $0.5269\pm0.0005$\\
        				&  						& ($-7.7\pm0.2$ min)\\
        $\phi_2$		& $0.0<\phi_2<1.0$ 		& $0.215\pm0.005$\\
        				&  						& ($5.0\pm0.7$ min)\\
        $\sigma$		& $\propto 1/\sigma$ 	& $0.0582\pm0.0004$\\
		\hline
	\end{tabular}
	\label{tab:model_lc_results}
\end{table}

As in the previous section, we find that the ``heating'' component arrives early by 7.7$\pm$0.2 minutes, while the ``ellipsoidal'' component arrives late by 5.0$\pm$0.7 minutes. We also find that the most appropriate error for the data is 0.0582$\pm$0.0004, which is exactly the same as the value of the covariance defined in Equation \ref{e:covariance_matrix} for a zero lag. This is higher than the expected photometric precision of the \textit{K2} mission, and is most likely related to variations occurring on a time scale shorter than 1~min.

\section{Short term variability}\label{sec:Flickering}

The model light curve was then subtracted from the data shown in Figure~\ref{fig:model_lc}. This residual light curve should then simply represent variability from the accretion disc alongside statistical noise, with all flares and orbital modulation removed. We computed the power spectrum (down to a frequency of 1 day$^{-1}$ to avoid the break which occurs at lower frequencies) of this residual light curve using the Lomb-Scargle periodogram method \citep{1976Ap&SS..39..447L,1982ApJ...263..835S}, and found that it was best described by a power law with index $\alpha=-(0.70\pm0.05)$. A random walk noise in the flux of the object should lead to a $f^{-2}$ behaviour in the power spectrum (\citealt{1995A&A...300..707T}; also called red noise), yet the power spectrum shown here is much closer to $f^{-1}$ (pink noise). There is no break to ``white'' measurement noise evident, implying that the intrinsic variations in the accretion disc emission are detected at all timescales down to $\lesssim1$~minute. The error in the power law fit to the power spectrum was found by generating several fake power spectra with known spectral indices using the algorithm described in \citet{1995A&A...300..707T} and implemented in \textsc{Stingray}\footnote{\textsc{Stingray} is a \textsc{Python} package for X-ray astronomy, and is available at \url{https://github.com/StingraySoftware/stingray}}. We then tested the accuracy of our fitting routines against these fake power spectra. 

This is significantly flatter than the power law indices of $-\left(1.2\pm0.1\right)$ and $-1.12\pm0.01$ found by \cite{2015MNRAS.453.3461S} and  \citetalias{2018arXiv180104736P} respectively. However, we find that we can reproduce their results if we do not remove the orbital modulation from the light curve before generating the power spectrum and calculating the power law index. This leads us to believe that leaving the orbital modulation in the light curve affects the measurement of the pink noise of the spectrum, and it should be removed in order to get an accurate estimate of the shape of the power spectrum caused by flickering within the system. While this explains the discrepancy between our results and those of \citetalias{2018arXiv180104736P}, this is not where the discrepancy between our result and the value found by \cite{2015MNRAS.453.3461S} arises, as they removed the orbital modulation before computing the spectral index.

Finally, we investigated the power spectrum of the data once the orbital modulation had been removed, but with flares reintroduced (as opposed to the above, where the flares were removed before computing the power spectra). The best fit power law to this spectrum was $-1.11\pm0.05$, which is also consistent with the result from \citetalias{2018arXiv180104736P}. This is also likely where the discrepancy between our result and the result of \cite{2015MNRAS.453.3461S} arises, as they included flares when calculating their power spectra.

\begin{figure}
	\begin{centering}
    \includegraphics[width=\columnwidth]{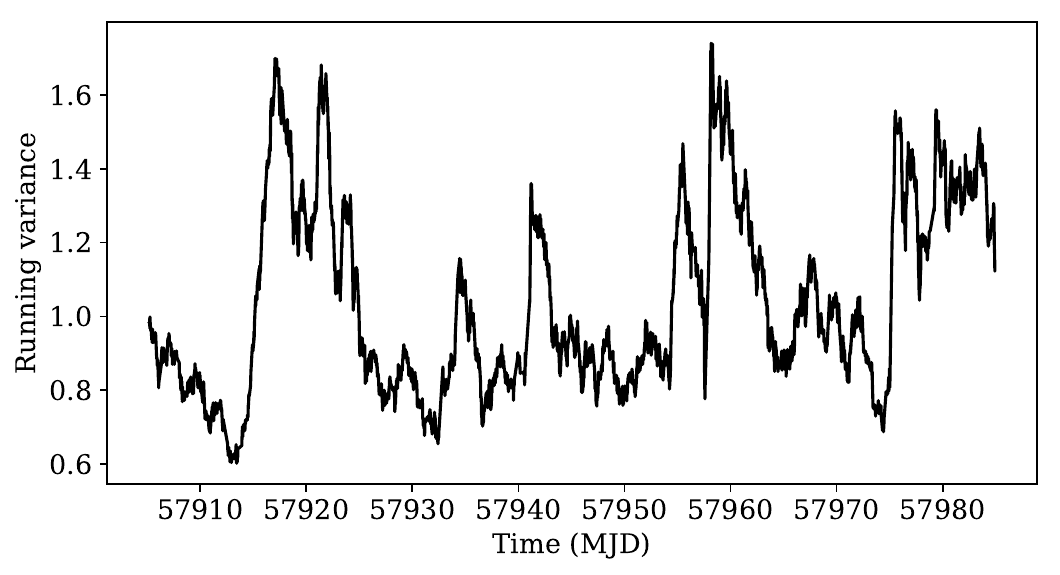}
    \caption{\label{fig:running_variance} Running variance of the whitened, flare subtracted residuals of Section~\ref{sec:Timing}, evaluated over 1-day windows, showing periods of enhanced variability. }
    \end{centering}
\end{figure}

The residual activity in the light curve, after subtracting flares and the orbital modulation, also varies significantly with time. This is evident in Figure~\ref{fig:running_variance}, which shows the running variance of the residuals obtained in Section~\ref{sec:Timing}, whitened by the averaged (stationary) covariance function. 

\begin{figure}
	\begin{center}
		\includegraphics[width=\columnwidth]{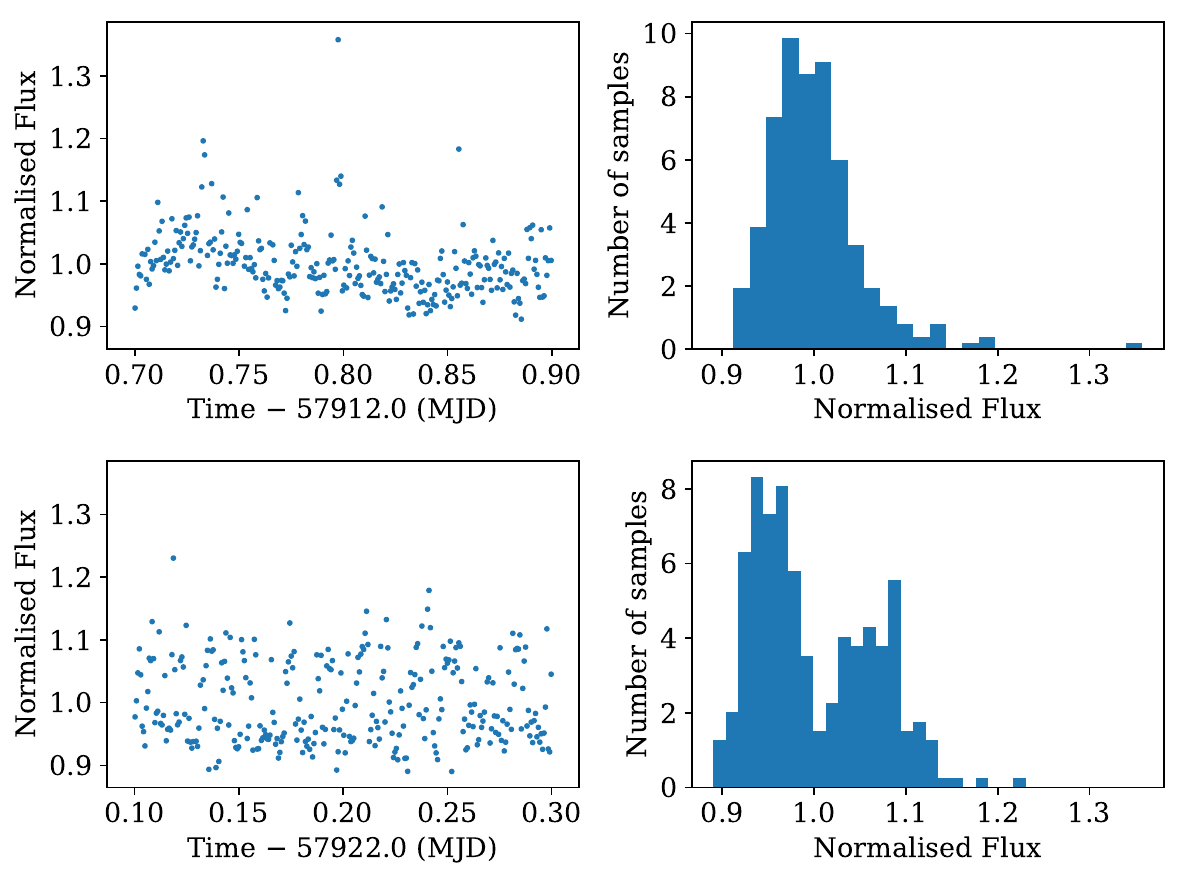}
		\caption{\label{fig:bimodal} Two equal-length time periods in which J1023 exhibits notably different variability modes. Left panels show the flux as a function of time after subtracting the orbital modulation and dividing by the mean value over the period shown. Right panels show histograms of these flux values. Upper panels show a period with ``flickering'', lower panels show a period in which J1023 switches between two distinct flux states over time scales of minutes, similar to behaviour observed by \citet{2015MNRAS.453.3461S}. }
	\end{center}
\end{figure}

Motivated by this, we inspected regions of the light curve where the variance was at its maximum and minimum. Figure~\ref{fig:bimodal} shows two such periods alongside histograms of the residual flux for both periods. Close to the time of maximum residual variance, the \textit{K2} light curve shows a similar bimodal flux distribution to that seen by \citet{2015MNRAS.453.3461S}. Due to the relatively long cadence of the observations relative to the duration of either of the individual modes, (some of the high and low modes shown in Figure~\ref{fig:bimodal} appear to only last for the duration of a single data point), this mode changing behaviour is not clearly resolved. Flickering and slower variations in the baseline also make this bimodal behaviour difficult to identify. However, we identify periods showing similar behaviour close to the other peaks in the residual variance, suggesting that the variability shown in Figure~\ref{fig:running_variance} may be caused by distinct activity modes.

\section{Flaring Activity} \label{sec:Flares}
\subsection{Flare selection method}
We now return to the original flattened light curve which contains both the orbital modulation and flares, but has been divided by the long term trend discussed in Section~\ref{sec:reduction}. After subtracting the model of orbital modulation (Section \ref{sec:Model}) from this flattened light curve, the remaining signal contains:
\begin{itemize}
	\item Correlated Gaussian noise with correlation matrix defined in Equation \ref{e:generalised_least_squares},
	\item unmodelled long-term and/or periodic modulations,
	\item flares.
\end{itemize}
Therefore, we shall define flares as contiguous sets of points that are unlikely to be produced by Gaussian noise. For a contiguous set of $n$ points, this condition is determined by $p$-value test on the $\chi^2$ determined from the residuals defined in Equation~\ref{e:generalised_least_squares} such that:
\begin{equation}
\chi^2 > \chi^2_{\mathrm{Max}}(n),
\end{equation} 
where  $\chi^2_{\mathrm{Max}}(n)$ is defined by the relation
\begin{equation}
\label{eq:flarecondition}
P(\chi^2 > \chi^2_{\mathrm{Max}}(n)) = \gamma(n/2, \chi^2_{\mathrm{Max}}(n)) = 0.001 \, ,
\end{equation}
where $P(\chi^2 > \chi^2_{\mathrm{Max}}(n))$ is the probability that $\chi^2$ be larger than  $\chi^2_{\mathrm{Max}}(n))$, and the threshold chosen to be 0.1\%.

We then added the additional condition that a contiguous segment marked as a flare must contain only one peak above the 3-sigma level of the uncorrelated noise ($\sigma = 0.058$). This condition prevents selecting unmodelled periodic or long-term modulations.

The flare selection algorithm first selects all 1-point sets according to the two criteria defined above, then removes them from the data set, and iterates with the 2-point sets and so on. Removing the already flagged points ensures that further investigation on a larger set length is not biased by points already known to deviate, and is therefore conservative in estimating duration of flares. The analysis is performed using a sliding window for each set length in order to catch every possible flare start time. We found that this algorithm tends to converge after a few tens of iterations and hence we stopped the analysis at a 40-point window. The resulting ensemble of flagged points are then merged into contiguous sets, which are potentially much longer than the size of our largest window. If a merged contiguous set contains more than one peak above the 3-sigma threshold (i.e. excursions above threshold separated by lower intervals), then the set is divided into individual flares where the lowest points between peaks mark the boundaries.

One advantage of our flare-selection method is that it avoids retaining too many false positives. Indeed, we find that choosing a different p-value threshold does not increase the number of points marked as flares by the amount expected from Gaussian noise. In addition, our method is also able to catch the tails of the flares that extend below the 3-sigma threshold, by virtue of the fact that a monotonously decreasing/increasing set of points is very unlikely to be generated by noise.

\subsection{Occurrence times}

\begin{figure*}
	\begin{center}
		\includegraphics[height=0.9\textheight]{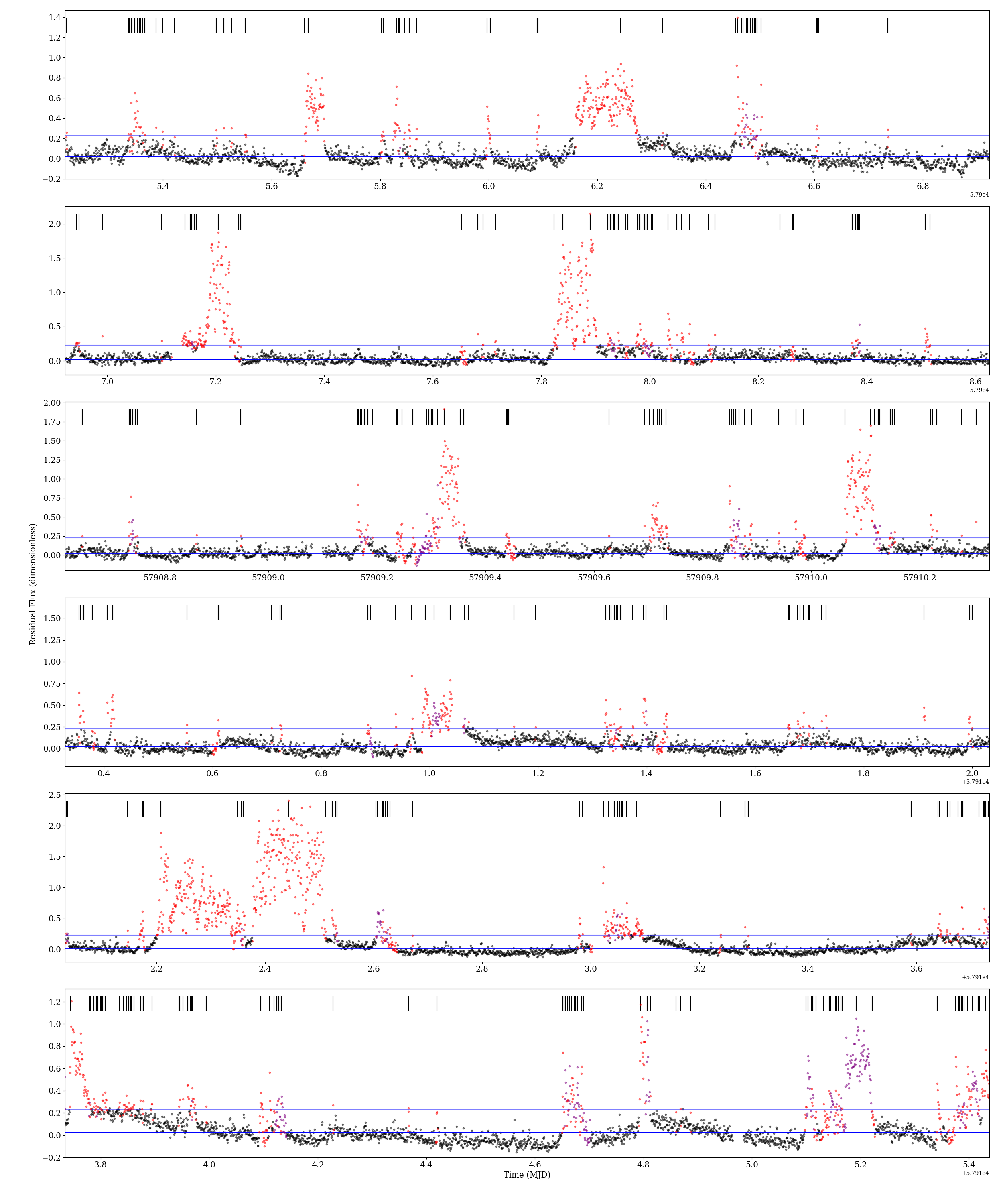}
		\caption{\label{fig:flare_light_curve} First 15,000 points (approximatively the first 14 days) of the flattened light curve with flares peaking above $3\sigma$, where $\sigma$ is the standard deviation of the base line (black dots). The flares are shown in alternate purple and red to allow distinguishing contiguous flares, and upper ticks give the position of each flare peak. The thick blue horizontal line shows the mean of the base line, while the thin blue horizontal lines show the $3\sigma$ level from the mean. (The full light curve is available as a online supplement) }
	\end{center}
\end{figure*}

From our previous analysis we found a total of 2880 individual flares, which together imply that the system spent $\sim 22\%$ of the time in a ``flaring'' state during the \textit{Kepler} \textit{K2} mission (see Figure~\ref{fig:flare_light_curve}). The number of flares we detect is significantly larger than the $\sim 170$ reported by  \citetalias{2018arXiv180104736P}. This could be due to differences in our definition of what constitutes a flare, which in particular involves splitting contiguous sets of points into sets of individual flares (with a single peak each). We will return to this at the end of the following section.

The standard deviation of the light curve from which flares were removed yields a value of 0.068, which is larger than the initial 0.058 found in Section~\ref{sec:Model}. Even taking into account this discrepancy, the distribution of the points of the flareless lightcurve shows an excess at large amplitudes compared to a Gaussian distribution. These two points strengthen our initial finding from Section~\ref{sec:Flickering} stating that the light curve contains a combination of low-level variability and bimodal states which contribute to increasing the global standard deviation, and introduce non-Gaussianity. 

\begin{figure}
	\begin{center}
		\includegraphics[width=0.5\textwidth]{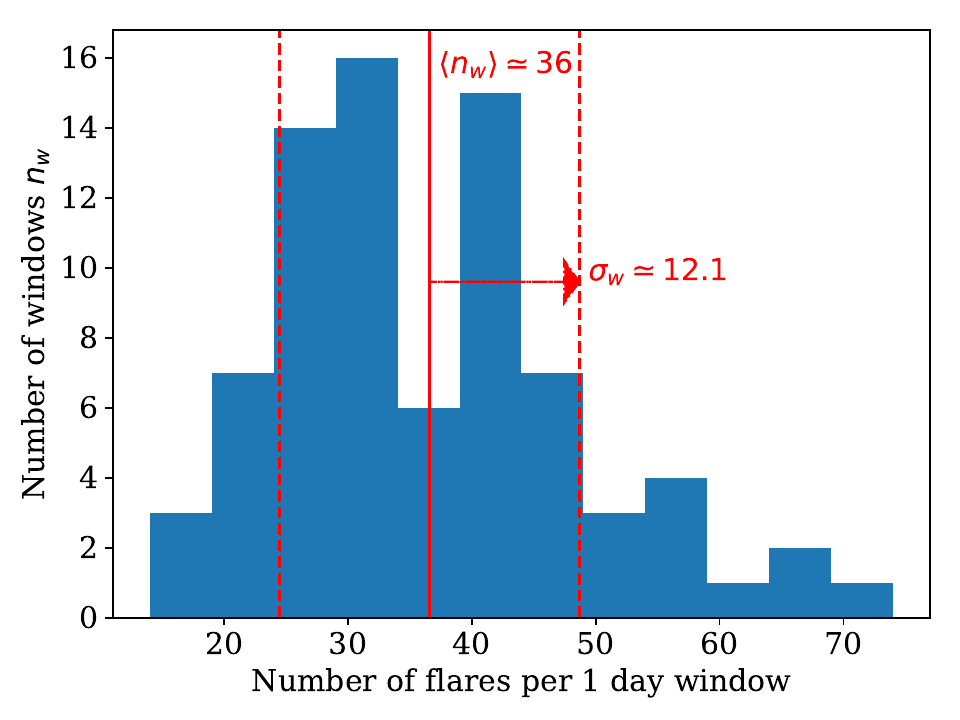}
		\caption{\label{fig:flare_freq} Distribution of the number of 1.0 day windows $n_w$ containing each number of flares. The number of flares is split into 10 bins. No satisfactory fit could be made based on a Gaussian or Poisson distribution. The average of the distribution $\left<n_w\right>$ is represented as a solid red line, and the $\pm$ one standard deviation $\sigma_w$ as dashed red lines.}
	\end{center}
\end{figure}

\begin{figure}
	\begin{center}
		\includegraphics[width=0.5\textwidth]{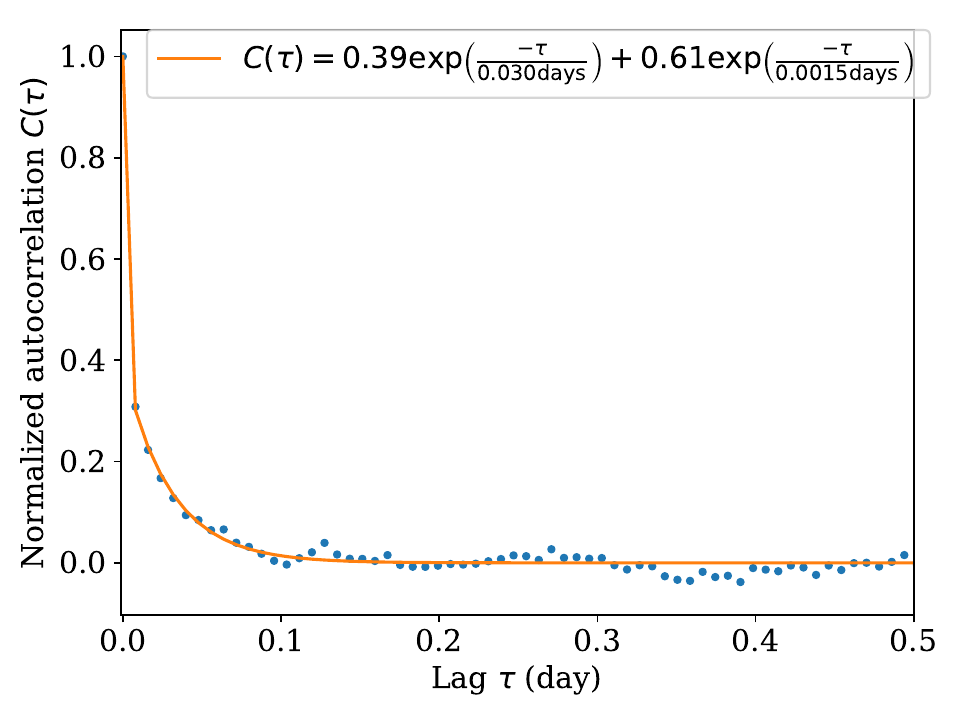}
		\caption{\label{fig:flare_autocor} Autocorrelation function of the number of flares occurring in 11.5min windows. The function is normalized by the variance of the distribution. A good fit  is obtained by a double exponential function (orange solid line). }
	\end{center}
\end{figure}

\begin{figure}
	\begin{center}
		\includegraphics[width=0.5\textwidth]{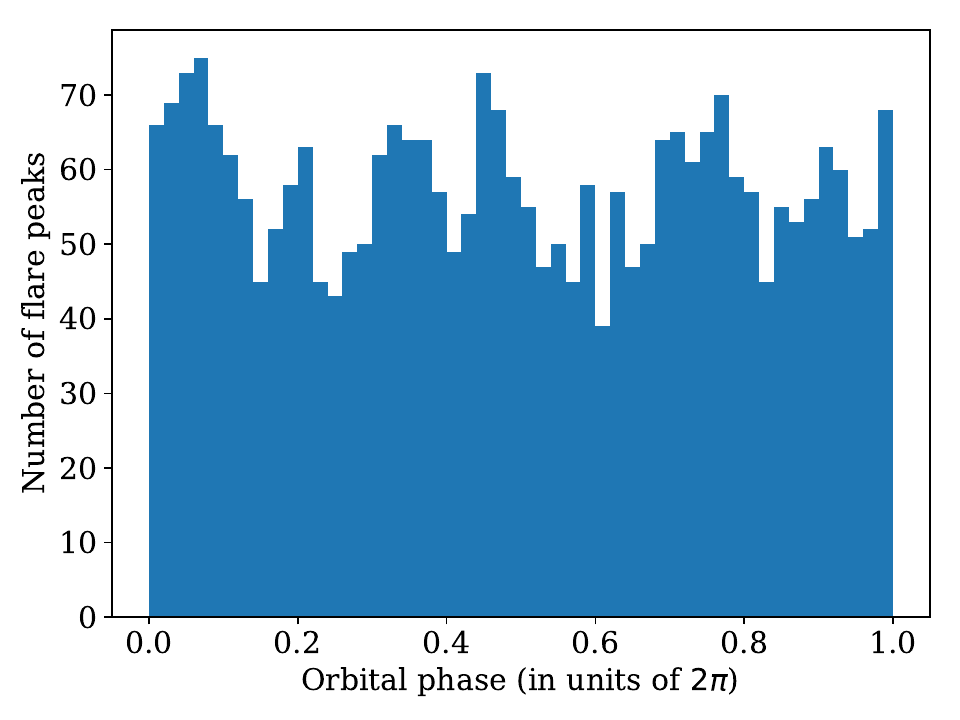}
		\caption{\label{fig:orbdistro} Distribution of the orbital phases of the peaks of the flares. Although the scatter in the number of flares is significant, we attribute this to clustering of flare events rather than an indication of an orbital dependence on flare rates.}
    \end{center}
\end{figure}

Figure~\ref{fig:flare_freq} shows the distribution of the number of flares occurring in a window of 1 day. On average, 36 flares per day have been detected, with standard deviation of 12 flares. The distribution is noticeably asymmetric and does not match a Gaussian or a Poisson distribution as naively expected from many typical random processes.

This hypothesis is reinforced by the autocorrelation function of the flares, Figure~\ref{fig:flare_autocor}, which shows correlations between flares over an exponential-decay time scale of about 43 min ($0.030$ days). 

As can be seen in Figure~\ref{fig:flare_light_curve}, there are contiguous successions of flares as the light curve goes rapidly (sometimes within one data point i.e. 1~min) below and above the 3-sigma threshold around the most intense flares. Whether these successions of flares should be considered as single long events, or studied as a separate subpopulation is not clear. When  excluding  these successions from the sample, the main difference to our analysis is that the flare frequency distribution (Figure~\ref{fig:flare_freq}) monotonously decreases with the number of flares per day.

We also investigated a possible link between the flaring activity and other parameters such as average luminosity of the baseline (Figure~\ref{fig:full_lc}) or the increase in local variability (Figure~\ref{fig:running_variance}). None of the previous quantities seem to present any correlation.

\newcommand{\Porb}{P_\mathrm{orb}}
\newcommand{\norb}{n_\mathrm{orb}}
\subsection{Flaring activity's dependence on orbital phase}
Our analysis did not reveal any clear correlation between the orbital phase and the flaring rate. Figure~\ref{fig:orbdistro} shows a histogram of the orbital phases of the peaks of the flares detected by our algorithm. While this histogram deviates from uniformity, there is no clear modulation at the orbital period as would be expected if flares were preferentially produced when the heated face of the companion is visible (c.f. the hints of orbital phase dependence of optical and X-ray flares from the black widow PSR~J1311$-$3430 by \citealt{2017ApJ...850..100A}). 

Instead, we attribute the apparent non-uniformity of flares' orbital phase distribution to the fact that flares do not appear to be independent events with a constant arrival rate, but are correlated with one another, as evident from Figure~\ref{fig:flare_autocor}. As a result, flares will tend to be ``clustered'' over short time scales, leading to excesses at random orbital phases. In other words, neighbouring orbital phase bins may not be independent, and their numbers do not follow Poisson statistics. As such, typical periodicity tests (e.g. the $H$-test \citet{1989A&A...221..180D}) which assume independence of events will over-estimate the significance of the non-uniformity of the flares' orbital phase distribution. Indeed, the $H$-test for the orbital phases of flare peaks is significant at the 3.3$\sigma$ level, but finds most power at the third harmonic of the orbital frequency, making it hard to explain this with a true viewing-angle explanation. We also computed a Lomb Scargle periodogram (\citealt{1976Ap&SS..39..447L}; \citealt{1982ApJ...263..835S}) using only the data points attributed to flaring activity by our algorithm (rather than the flare peak times), and again found no evidence for periodicity at the orbital frequency.

This is a major difference between our analysis and that of \citetalias{2018arXiv180104736P} who find an enhanced flaring activity around phase 0.75. One difference between our analyses is that \citetalias{2018arXiv180104736P} define flares by an iterative sigma-clipping method. An additional difference is that we considered only the orbital phase of the peak of each flare while \citetalias{2018arXiv180104736P} consider every data point that lies above their (iterated) significance threshold. We were able to construct a flare distribution similar to that seen in \citetalias{2018arXiv180104736P}, but this required us to include every individual point that was identified as being part of a flare, a set of highly non-independent events since many form contiguous sets belonging to the same flaring episodes. Furthermore, this distribution peaked at phase ~0.4, almost out of phase with the excess observed in \citetalias{2018arXiv180104736P}.  We therefore conclude that an orbital phase dependence of the flare rate is unclear, and highly dependent on one's chosen definition of flares and the time at which they occur.

\subsection{Peak intensities, fluences and durations}

\begin{figure*}
	\begin{center}
		\includegraphics[width=1\textwidth]{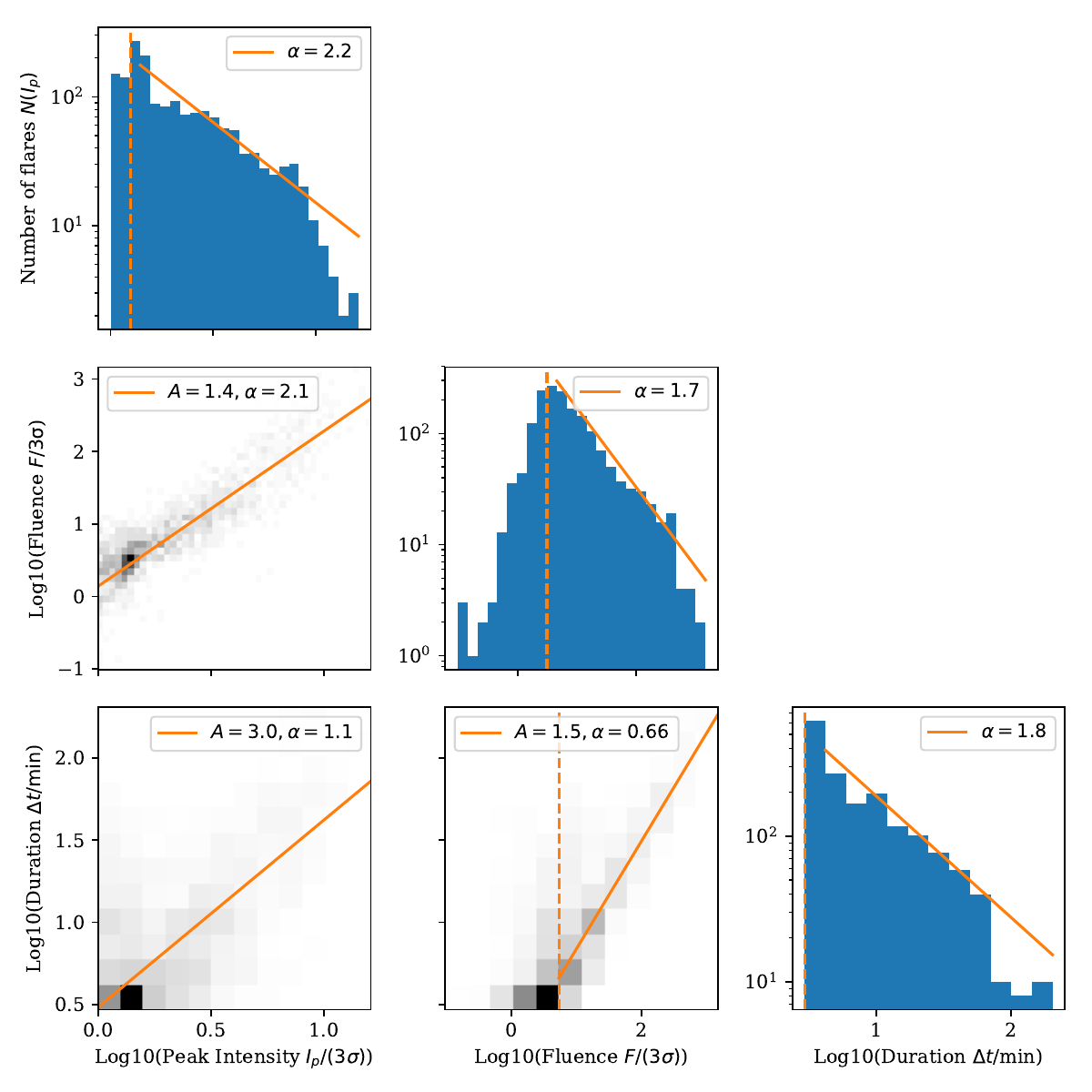}
		\caption{\label{fig:flarestatcorner}Corner plot of the peak intensity, fluence and duration of the flares. The three quantities are organised in rows and columns as in a matrix: on the diagonal are plotted the histograms of the corresponding flare quantities, and elsewhere the 2D covariance histograms. The peak intensity and the fluence are normalised by the threshold flux $3\sigma$ where $\sigma =0.058$. The cumulative distributions are fitted with power-law distributions  corresponding to the probability density function (PDF) $\frac{\mathrm{d}P}{\mathrm{d}x} = \left(\frac{x}{x_0}\right)^{-\alpha}$ where $x$ is the quantity under consideration (e.g. $I_p/(3\sigma)$ and $x_0$ is determined by the normalisation condition of the PDF. The PDFs are plotted as solid orange lines. Similarly, the correlations are fitted with laws of the type $\left<y\right>_{\mathrm{geom}} = A x^{\alpha}$ represented as a solid orange line. The lower limit of the fitting range is shown by the vertical dashed orange lines (when applicable). }
	\end{center}
\end{figure*}

The distribution of the intensities of the peaks of the flares, Figure~\ref{fig:flarestatcorner}, is well fit by a power-law with an exponential cutoff, with 90\% of the flares peaking between $3\sigma$ and $3.4(3\sigma)$, with a number of outliers extending up to $13 \sigma$. 

The monotonously decreasing shape of the peak intensity distribution  (Figure~\ref{fig:flarestatcorner}) suggests that it is only the high-intensity  tail of a more complete distribution. A possibility would be that the noise level is actually lower, but very frequent small flares (below 3-sigma) and the aforementioned optical mode switching actually widen the baseline. This idea is further re-enforced by the results of Section \ref{sec:Model} which showed that the scatter of individual data points was much higher than expected for pure noise.

The fluence $F$ of a flare is calculated as the sum of the flux of its individual data points. The fluence appears to be well correlated (Figure~\ref{fig:flarestatcorner}) with the peak intensity $I_p$ with the relation
\begin{equation}
\left<\frac{F}{3\sigma}\right>_{\mathrm{geom}} \simeq 1.4 \left(\frac{I_p}{3\sigma}\right)^{2.1},
\end{equation}
where $\left<\right>_{\mathrm{geom}}$ denotes the geometric mean. This relation (and others that follow) was obtained from a linear fit performed between the logarithm of the two quantities, hence the use of geometric mean instead of arithmetic mean.
The base durations of the flares, taken between the first and the last point flagged by our algorithm, is well fitted by a double exponential function. The flares last between 1 min (only one point) and 204 min for the longest. We note that the lower limit may be partially due to the time resolution of the data, but also to the inability of our algorithm to efficiently detect the tails of low amplitude flares, while the upper limit hints at what may be causing the difference in the number of detected flares between our results and \citetalias{2018arXiv180104736P}, as \citetalias{2018arXiv180104736P} report their longest flare to last 14 hours, suggesting we have split up their longest flare into multiple shorter events.

We investigated the relation between the flare peak intensities and their base duration (Figure~\ref{fig:flarestatcorner}) after first removing the flares lasting less than 3 minutes in order to limit the effect of the aforementioned bias. The result is that the average duration $\left<\Delta t \right>$ is almost proportional to the peak intensity,
\begin{equation}
\left<\Delta t \right>_{\mathrm{geom}}(I_p) \simeq 3.0 \left(\frac{I_p}{3\sigma}\right)^{1.1} \mathrm{min}.
\end{equation}
It is to be noticed that this relation gives a minimum mean duration of 3 min for a flare above the 3-sigma threshold, therefore suggesting that our algorithm does not easily detect the tails of the smallest flares. 
The relation between the (geometric) mean fluence and the base duration of the flares is also well correlated as expected from the previous relations between the peak intensity and the fluence base duration, as can be seen on Figure~\ref{fig:flarestatcorner}.

\section{Discussion}\label{sec:Discussion}
\subsection{The companion's orbital light curve}
The observed orbital light curve (Figure~\ref{fig:model_lc}) is remarkably similar to the light curve of J1023 presented in both Figure 7 of \citet{2004MNRAS.351.1015W} and Figure 4 of \citet{2005AJ....130..759T}, with all three light curves showing a similar asymmetry. However, the observations in \citet{2004MNRAS.351.1015W} and \citet{2005AJ....130..759T} were taken when the system was in a non-accreting state (we are tempted to call this the radio powered state but will not, as the system had not been detected as a pulsar at the time). We also tested whether or not the peak-to-peak amplitude of the light curves in both states were the same by converting the SAP flux to \textit{Kepler} magnitudes. The peak-to-peak amplitude due to orbital modulations in both states was close to 0.35 mag.

This suggests that the amount of heating the secondary star experiences in both states is approximately equal, which is surprising given the increase in gamma-rays from the NS in the accretion powered state by a factor $\sim 6$ found by \citet{2014ApJ...790...39S}. This indicates that gamma ray photons are possibly not the main source of irradiation heating the companion, thus leaving the pulsar wind (the dominant mechanism for angular momentum loss during the rotationally-powered pulsar state) as the primary contender. 

Furthermore, \citet{2016ApJ...830..122J} found that the neutron star's spin-down rate is $26.8$\% faster in the accreting state than in the rotationally-powered radio pulsar state. This increase is surprisingly small given the dramatic effects that an accretion disc is expected to have on the pulsar's magnetic field structure. The similarity in the heating profile observed here compared to previous observations during the radio pulsar state provides further evidence that the pulsar wind spin-down mechanism is not suppressed by the accretion disc encroaching into the pulsar's magnetosphere. 

The clear asymmetric component to the heating signal could be caused by a combination of asymmetric re-processing of the heating flux in a shock between pulsar and companion \citep{2016ApJ...828....7R}; heat redistribution on the surface of the irradiated companion introduced by its rotation (\citealt{1995MNRAS.275...31M}; \citealt{2000MNRAS.314..747S}); and/or by spots on the companion \citep{2016ApJ...833L..12V}. Understanding these asymmetries will be of key importance for explaining the nature of the mechanism heating the companion star in irradiated pulsar binary systems.

Additionally, thanks to the long period of continuous monitoring offered by the \textit{Kepler} data, we tentatively identify slight differences in the apparent periods of these components with respect to the known orbital period, suggesting that whatever causes the observed asymmetry may vary with time. The measured period offsets suggest that the asymmetry may have been increasing over the \textit{K2} observations, with the optical maximum drifting to earlier times compared to the known orbital phase. In principle, precession of the pulsar's ascending node  could lead to a similar effect. This precessing phenomenon has been observed as in other redback systems \citep[e.g.][]{2015ApJ...807...18P} as variations in their orbital periods, and is thought to be caused by their companions' varying gravitational quadrupole moments. However, the orbital period derivative measured by timing pulsations from J1023 \citep{2016ApJ...830..122J} would lead to a shift in the time of ascending node of less than one second over the \textit{K2} observations, and hence cannot explain the measured period offset of the heating component. Alternatively, this effect may be due to a change in the heat redistribution on the secondary star over the duration of the observations (which would alter the apparent fundamental frequency). 

For other redback systems which are not monitored over long time periods as closely as J1023 has been with \textit{Kepler}, these effects would manifest as markedly different light curves in widely separated observations.

A detailed understanding of the possible differences between the dynamical orbital phase and the apparent phase of the companion stars optical light curve may be crucial for searches for gamma-ray pulsations from black-widow or redback candidate systems \citep[e.g.][]{2014ApJ...793L..20R,2015ApJ...814...88S}. These searches rely on extremely precise knowledge of the putative pulsar's orbital phase, derived from the light curve of the companion star, to account for the Doppler shifting of the pulsar's pulsations due to its projected orbital velocity \citep{2012Sci...338.1314P}. Understanding that the apparent optical light curve may be shifted by several minutes when compared to the true time of ascending node, and that this shift may change with time, will guide the parameter space of orbital phases and periods which must be searched.

\subsection{Stochastic Variability}
In Section~\ref{sec:Flickering}, we found the power spectrum of the \textit{K2} data to have a power-law shape with a spectral index of $\alpha=-(0.70\pm0.05)$. This spectral index is very close to the $f^{-1}$ (pink noise) behaviour expected from flickering in the light curve \citep{1987Natur.325..694L}. This flickering spectrum extends to the Nyquist frequency without any evidence for a break to ``white'' noise. Such a spectral break is only seen at frequencies above $1$~Hz by \citet{2017NatAs...1..854A}, suggesting that this flickering occurs at least down to $\sim1$-second time scales.

The \textit{K2} optical light curve of J1023 also shows similar properties to the X-ray light curve of the source during the accretion state - that is, it has two modes (a low and high mode) alongside flickering and flares. The nature of the mode switching between both high and low modes in the optical light curve has been studied previously \citep{2015MNRAS.453.3461S}. \citet{2015ApJ...806..148B} reported a non-detection of bimodal behaviour in their optical observations. Our results here compliment both of these results, as we clearly see periods when the mode switching behaviour is detectable and undetectable in the optical light curve. We do note that \citetalias{2018arXiv180104736P} reported a non-detection of optical mode switching in their paper on the same data. A reason for this is that \citetalias{2018arXiv180104736P} only considered the dataset as a whole or in 2 day intervals when looking for bimodal behaviour, while we find it occurring on much shorter timescales (the data shown in Figure~\ref{fig:bimodal} represent one of the clearest examples of bimodality in the light curve, which only covers $\sim0.2$~d before being interrupted by a flare). Simultaneous X-ray and optical observations would be required to determine whether or not the optical high and low modes are correlated with the observed X-ray mode switching. 

One of the more unexpected results from analysing the flares is that the flare rate did not depend on the baseline flux of the system. This is surprising, as if the increase in baseline flux of the system around the times highlighted in Figure~\ref{fig:full_lc} are due to an increase in mass transfer rate, and the flares are produced by a build up of material at the inner radius of the disc, then an increase in the mass transfer rate should lead to more frequent and/or more powerful flares during these periods. We note the de-trending of the light curve cannot be responsible for the lack of correlation as raw data in Figure~\ref{fig:full_lc} does not show any link between baseline flux and flaring activity either.

\subsection{Self-organised-criticality analysis}

\begin{figure}
	\begin{center}
		\includegraphics[width=0.5\textwidth]{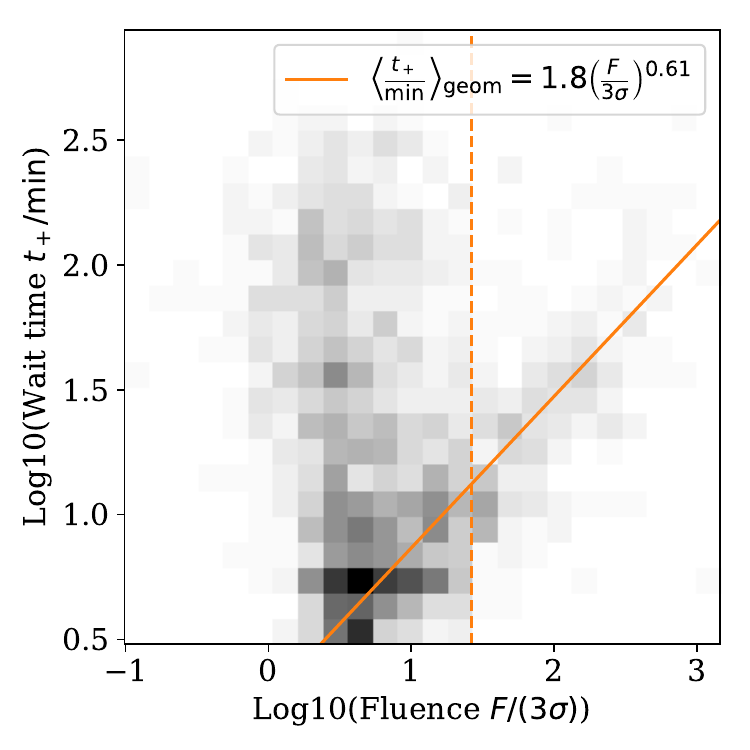}
		\caption{\label{fig:waittimevsflu} Distribution of the flares as a function of their fluences and of the time elapsed before the next flare (wait time $t_+$). In both cases, a fit of the mean wait time as a function of the fluence is made on the bins right to the vertical dashed orange line, and the result shown as the solid orange line.}
	\end{center}
\end{figure}

\begin{figure*}
	\begin{center}
		\includegraphics[width=1\textwidth]{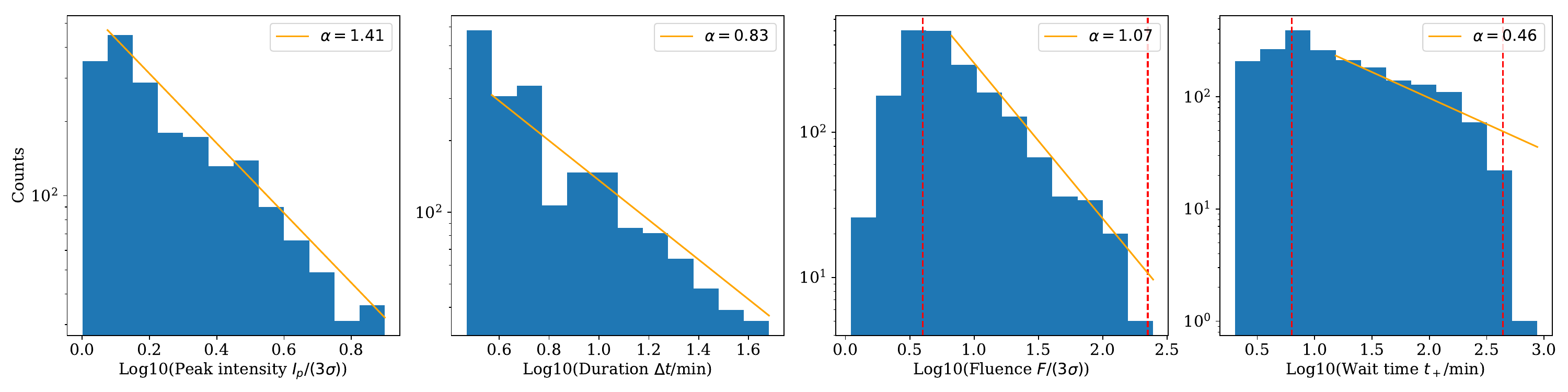}
		\caption{\label{fig:socanalysis} Distribution of the peak intensity, duration, fluence and wait time $t_+$ for a subsample of flares with peak intensities in the range $ \frac{I_p}{3\sigma} \leq 7.9$, durations in the range $2.8 \mathrm{min} < \Delta t \leq 49\mathrm{min} $  totalling 1198 flares. This range is chosen to select power-law behaviours of all these quantities, in order to test the self-organised criticality hypothesis (see text). Orange lines represent the power-law fit on each distribution, characterised by a the index $\alpha$. The fits were made using the integral of the distribution (as in Figure~\ref{fig:flarestatcorner}). In the case of the fluence and the wait time, fitting interval had to be restricted to the area between the two vertical dashed red lines.}
	\end{center}
\end{figure*}

We investigated the possibility that the flares might be the result of a self-organised criticality (SOC) mechanism (see \citet{aschwanden_25_2016} for a review). In this kind of system, the energy provided by an external driver accumulates in the system which gradually reaches a state locally close to instability in multiple regions. The energy is then violently released when these instabilities are reached. The instabilities themselves are triggered in an apparently random way, and are subsequently regenerated under the action of the driver.This kind of process is thought to occur in many different fields, particularly earthquakes but also Solar flares and X-ray flares of gamma-ray-burst afterglows (see e.g. \citet{wang_self-organized_2013}). 

One expected property is that the wait time is correlated with the amount of energy released in a flare. The wait time is defined as the time elapsed before the next flare (noted here $t_+$, we also consider the time since the previous flare $t_-$ for comparison), with the flares being stamped with their peak time. We show the distribution of flares with respect to the fluence and the wait time on Figure~\ref{fig:waittimevsflu}. A correlation is visible for the higher fluence events, while there is no correlation for the lower fluence events. This separation in two branches can be explained, at least partly, by the fact that the fluence of the lower flares is not well estimated by our algorithm as part of the tails of these flares might be missing. It is also possible that the wait times associated with the smaller flares are primarily determined by occurrences of flares peaking below our 3-sigma threshold, or that different subpopulations of flares exist. Interestingly, although the same two-branch correlation is visible for both $t_-$ and $t_+$, the high-fluence correlation is clearer in the case of $t_+$.

The SOC theory predicts precise power-law indices for several quantities depending on the dimension of the problem \citep{aschwanden_25_2016}. 
We investigated the potential power-law behaviour of the distributions of peak intensities, durations, fluences and wait times (Figure~\ref{fig:socanalysis}). These distributions are plotted from a subsample of flares from which low peak-intensity and low-duration flares have been removed in order to avoid ill-defined durations and fluences. The flares representing the higher end of these distributions have been removed to avoid very low statistics. All these quantities can be well fit by power laws at least on a large interval.
In the case of the duration (Figure~\ref{fig:socanalysis}), the SOC theory suggests the power-law index should equal the dimensionality of the system. Here the measured index is $\alpha = 0.83$ suggesting a mono-dimensional system. Assuming a strictly one-dimensional system, the theory predicts an index of 1 for both the peak-intensity and fluence distributions while we find respectively 1.41 and 1.07. Note that we consider here that the peak intensity is proportional to the maximum flux of the flare, which is the case if the peak is not occurring on a much shorter time scale than the integration time of the data points. The prediction for the wait-time depends heavily on the hypothesis made on the stochastic distribution underlying the flares: a simple Poisson distribution predicts an exponentially decaying wait-time distribution, while different non-stationary Poisson processes (with mean occurrence rates which vary with time) predicts different results, often with a power-law tail at large wait times \citep{aschwanden_25_2016}. Here, the best power-law fit gives an index of 0.47, but we do not draw any interpretation from it.

We also remark that the power-law index of the Fourier spectrum of the light curve with flares but without orbital modulation is $-1.11 \pm 0.05$ while without flares it is only  $-0.70 \pm 0.05$ (see section \ref{sec:Flickering}). The pink noise (power-law index of $-1$) is one of the main characteristics of SOC processes, as this theory was introduced to explain the ubiquitous presence in natural processes of this type of noise \citep{bak_self-organized_1987}. Therefore, the fact that the index is closest to -1 with flares included suggests that these are only the larger-amplitude tail of an overall SOC signal. This is further re-enforced by the previously mentioned fact that the distribution of peak intensities behaves like the upper tail of a broader distribution. 

\subsubsection{A comparison with low accretion rate CVs}
The most obvious self organised criticality mechanism to compare J1023 during its accretion phase with is the SOC  possibly present in low accretion rate cataclysmic variables (e.g. \citealt{1998A&A...337..962D}). Low accretion rate non-magnetic CVs often undergo brightening events called dwarf novae during which the systems brightness increases by 2-7 magnitudes. Dwarf novae events are suspected to arise due to a combination of an instability in the accretion disc (disc instability model, or DIM; \citealt{1974PASJ...26..429O}) and a variable mass accretion rate. At low accretion rates, the discs are cool and stable. Once the discs reach a critical density, the viscosity of the disc changes, and material begins to move rapidly inward through the disc, increasing the temperature of the disc while rapidly transferring material on to the white dwarf. This continues until the density of the disc drops below the critical value, and the disc becomes stable again. For a review, see \citet{2001NewAR..45..449L}.

While the mechanisms between the outbursts in dwarf novae and the flares in J1023 are very different, future work should be done on investigating the similarities between these systems, especially since the driving force behind dwarf novae and the flares in J1023 is accretion.

\section{Conclusions}
The \textit{Kepler K2} light curve of J1023 has revealed more about the nature of tMSPs during their accreting state.

In particular, we have found that the amplitude of the underlying orbital modulation in the accretion state is similar to that in the radio powered state, while the asymmetry in the light curve is less obvious, but still detectable. The similarity between the light curves in these two distinct states suggests that the mechanism by which the neutron star heats the companion remains active and of similar magnitude in the accreting state, despite the large increase in the system's high-energy emissions \citep{2014ApJ...790...39S} and differing torques acting on the neutron star \citep{2016ApJ...830..122J}.

Secondly, we have found that the fractional time J1023 spends in an optical flaring state appears to be much higher than the system spent in the X-ray flaring state in the observations presented in \citet{2015ApJ...806..148B}. This either suggests that previous studies have underestimated their number (which is likely due to the much shorter baselines of their observations) or that the system was flaring more during the \textit{K2} observations than in previous years. This could be due to a higher mass transfer rate or more subtle changes in the disc, but without a definitive answer for where the flares are produced, it is impossible to say for certain what changes would give rise to a higher flare rate.

Finally, our analysis suggests that the SOC theory for the optical flares qualitatively, if not quantitatively, agrees with the predictions for a one-dimensional SOC theory: peak intensity, duration, and fluence show appropriate power-law behaviours, the wait time is correlated with the fluence as expected, and the light curve exhibits a marked pink noise. 

Continued optical, X-ray, radio, and gamma-ray observations of J1023 are highly encouraged, as not only might they help answer where the optical and X-ray flares are produced, and if the heating component of the orbital modulation changes, but there is also the chance that J1023 will transition back into another radio powered state, in which case new data will be invaluable in understanding the transitions within these systems.

\section*{Acknowledgements}

We would like to thank the anonymous referee for their comments which improved this paper. R.P.B., C.J.C. and G.V. acknowledge support from the ERC under the European Union's Horizon 2020 research and innovation programme (grant agreement No. 715051; Spiders). M.R.K is funded through a Newton International Fellowship provided by the Royal Society. M.R.K. would like to thank Paul Callanan for useful discussions. This paper includes data collected by the \textit{Kepler} mission. Funding for the \textit{Kepler} mission is provided by the NASA Science Mission directorate. Some/all of the data presented in this paper were obtained from the Mikulski Archive for Space Telescopes (MAST). STScI is operated by the Association of Universities for Research in Astronomy, Inc., under NASA contract NAS5-26555. Support for MAST for non-HST data is provided by the NASA Office of Space Science via grant NNX09AF08G and by other grants and contracts. This research made use of Astropy, a community-developed core Python package for Astronomy \citep{2013A&A...558A..33A}. This work also made use of SciPy \citep{scipy}.





\bibliographystyle{mnras}
\bibliography{j1023_bib} 







\bsp	
\label{lastpage}
\end{document}